\documentclass[usenatbib]{mnras}

\usepackage{newtxtext,newtxmath}

\usepackage[T1]{fontenc}
\usepackage{ae,aecompl}
\usepackage{times}

\usepackage{graphicx}	
\usepackage{amsmath}	
\usepackage{amssymb}	
\usepackage{bm}
\usepackage{calc}
\usepackage{color}
\usepackage{url}
\usepackage{array}
\usepackage{tabularx}
\usepackage{multirow}

\def\gsim{~\rlap{$>$}{\lower 1.0ex\hbox{$\sim$}}}

\def\simpropto{\lower.2ex\hbox{$\; \buildrel \propto \over \sim \;$}}
\def\ltsim{\lower.5ex\hbox{$\; \buildrel < \over \sim \;$}}
\def\gtsim{\lower.5ex\hbox{$\; \buildrel > \over \sim \;$}}
\def\ltsim{\lower.5ex\hbox{$\; \buildrel < \over \sim \;$}}
\def\gtsim{\lower.5ex\hbox{$\; \buildrel > \over \sim \;$}}

\def\prd{Physical Review D.}

\def\pmb#1{\setbox0=\hbox{#1}%
\kern-.025em\copy0\kern-\wd0
\kern.05em\copy0\kern-\wd0
\kern-.025em\raise.0433em\box0}

\renewcommand{\theenumi}{(\alph{enumi})}

\def\aj{AJ}

\def\simlt{\lower.5ex\hbox{$\; \buildrel < \over \sim \;$}}
\def\simgt{\lower.5ex\hbox{$\; \buildrel > \over \sim \;$}}

\newcommand{\beq}{\begin{equation}}
\newcommand{\eeq}{\end{equation}}
\def\beqa{\begin{eqnarray}}
\def\eeqa{\end{eqnarray}}
\def\fixit#1{}

\def\apj{ApJ}
\def\jcap{JCAP}

\def\apjs{ApJ. S}

\def\physrep{Physics Reports}
\def\apjl{ApJL}
\def\mnras{MNRAS}
\def\aap{A\&A}

\title[Probing velocity-density correlations with galaxy luminosities]{Probing cosmic velocity-density correlations\\ with galaxy luminosity modulations}

\author[M. Feix]{Martin Feix$^{1,2}$\thanks{E-mail: feix@uni-heidelberg.de}
\\
$^{1}$ {Zentrum f\"{u}r Astronomie der Universit\"{a}t Heidelberg, Institut f\"{u}r Theoretische Astrophysik, 69120 Heidelberg, Germany}\\
$^{2}$ {Institut f\"{u}r Kernphysik, Karlsruher Institut f\"{u}r Technologie, 76344 Eggenstein-Leopoldshafen, Germany}
}

\date{Accepted XXX. Received YYY; in original form ZZZ}

\pubyear{2018}

\begin{document}
\label{firstpage}
\pagerange{\pageref{firstpage}--\pageref{lastpage}}
\maketitle

\begin{abstract}
We study the possibility of using correlations between spatial modulations in the observed luminosity distribution of galaxies and
the underlying density field as a cosmological probe. Considering redshift ranges, where magnification effects due to gravitational
lensing may be neglected, we argue that the dipole part of such luminosity-density correlations traces the corresponding velocity-density
signal which may thus be measured from a given galaxy redshift catalogue. Assuming an SDSS-like survey with mean density $\overline{n}=0.01
(h^{-1}$ Mpc)$^{-3}$ and effective volume $V_{\rm eff}=0.2(h^{-1}$ Gpc)$^{3}$ at a fiducial redshift $z=0.1$, we estimate that the
velocity-density correlation function can be constrained with high signal-to-noise ratio $\gtrsim 10$ on scales 10--100 Mpc. Similar
conclusions apply to the monopole which is sensitive to the environmental dependence of galaxy luminosities and relevant to models of
galaxy formation. 
\end{abstract}




\section{Introduction}
\label{section1}
Within the standard cosmological paradigm, the process of structure formation is connected to matter flows that exhibit a coherent pattern on large scales.
Since the motions of astronomical objects such as galaxies are believed to reliably trace these matter flows, their peculiar velocities, defined relative to
the uniform expansion of the cosmic background, provide a sensitive probe of the underlying cosmological model \citep[e.g.,][]{Strauss1995}.

At cosmologically relevant distances, however, observations can only probe total velocity components along the line of sight that are difficult to measure
because surveys only provide redshifts, but not actual distances. Typically, these are estimated by resorting to empirical, redshift-independent distance indicators
that exploit established relations between observable intrinsic properties of galaxies or other astronomical objects \citep{TF77, Djorgovski1987}. Combining the inferred
distances with redshift data then yields estimates of the radial peculiar velocity field sampled at the corresponding object positions.

Traditionally, peculiar velocity surveys of this kind have been recognized as a valuable asset since they offer \emph{direct} constraints on the peculiar velocity
field of galaxies. Most notably, this includes measurements of the cosmic bulk flow, i.e. the volume average of the peculiar velocity field, that can be tested
against expectations of different cosmological models \citep[e.g.,][]{Nusser2011, Turnbull2012}. The inferred peculiar velocities may also be compared to
the large-scale velocity field independently predicted from the observed galaxy distribution using gravitational instability theory, allowing for estimates of
the growth rate of density fluctuations \citep[e.g.,][]{Erdo06, DN10, Hudson2012}. Unlike the clustering analysis of redshift-space distortions \citep[RSDs;][]{k87, Percival2009},
this latter approach is less affected by cosmic variance as it involves the ratio of correlated fields. The uncertainties mainly arise from the precision of the
peculiar velocity estimates themselves. In particular, peculiar velocity surveys allow for a simultaneous analysis of multiple tracers, i.e. the galaxy density and
the peculiar velocity fields, whose cross-correlations add independent information that leads to tighter cosmological constraints \citep{Nusser2017, Adams2017} and
is also important to the modelling of RSDs \citep{Koda2014}.

Due to observational challenges, however, peculiar velocity surveys suffer from complicated selection functions, sparseness, relatively small galaxy numbers in comparison
to galaxy redshift catalogues, and errors that increase rapidly with redshift. This limits reliable studies of the cosmic peculiar velocity field to the local Universe
\citep[but also see][]{Hellwing2017, Hellwing2018}. Current datasets like Cosmicflows-3 \citep{Tully2016} contain on the order of $10^4$ objects out to distances of about
100--150 Mpc. Next-generation surveys such as TAIPAN \citep{daCunha2017} and all-sky HI radio observations \citep[e.g., WALLABY;][]{Duffy2012} will extend to around twice
larger distances, increasing the number of objects by roughly an order of magnitude.

Recently, an alternative methodology towards constraining the large-scale peculiar velocity field from galaxy redshift surveys has been proposed. In contrast to
the analysis of peculiar velocity catalogues, this approach is not limited to small redshifts and does not rely on the use of traditional distance indicators. The method
uses the fact that galaxy peculiar motions affect luminosity estimates derived from measured redshifts which are used as distance proxies. Therefore, spatial modulations in the
observed distribution of galaxy luminosities can be exploited to place bounds on the underlying large-scale galaxy velocity field. Dating back to the work of \cite{TYS},
key to this luminosity-based approach is that peculiar velocities increase the scatter of (absolute) galaxy magnitude estimates about their true values, i.e. galaxies
generally appear brighter or dimmer than they would be if their redshifts accurately reflected the correct distances. In the case of spatially coherent motions, the
effect is systematic and depends on the position of galaxies in the observed survey volume.
With the advent of larger galaxy samples and improved photometry in current survey data, this method has started to become interesting for cosmological applications.
For example, it has been adopted to independently measure bulk flows \citep{Nusser2011, Branchini2012, Feix2014} and the cosmic growth rate \citep{Nusser2012, Feix2015,
Feix2017} out to redshifts $z\sim 0.1$.

Applications using galaxy luminosity modulations are not limited to one-point statistics associated with the luminosity distribution of galaxies. Tracing the radial
peculiar velocity field of galaxies, these modulations may also be studied in terms of spatial correlations that can be easily computed from the data and do not require
any smoothing or pre-modelling of the signal. Previous analyses have mostly focused on the correlation function of luminosity distances \citep[e.g.,][]{Bonvin2006, Hui2006,
Nusser2013, Biern2017}. The main purpose of this work is to motivate measurements of luminosity-density correlations as a probe of cosmic large-scale velocity-density
correlations beyond redshifts $z\sim 0$, and to provide a first assessment of such an approach.

The paper is structured as follows: in section \ref{section21}, we consider the various contributions to the large-scale luminosity signal and its correlation with
the underlying density field. Assuming Gaussian statistics and the distant observer approximation, we then discuss how well the sought cosmological signal could
be extracted and provide a first quantification of its statistical significance in sections \ref{section22} and \ref{section23}. Finally, we conclude in section
\ref{section3}.

\section{Methodology}
\label{section2}
\subsection{Basic considerations}
\label{section21}
Consider a galaxy redshift catalogue with measured apparent magnitudes $m_{j}$, angular positions $\hat{\bm{r}}_{j}$, and spectroscopic redshifts $z_{j}$, centred
around an effective redshift $z_{\rm eff}$. For each galaxy in the sample, we may estimate the luminosity modulation
\begin{equation}
\Delta M_{j} = M_{j} - \overline{M},
\label{eq:211}
\end{equation}
where $M_{j}$ is the observed absolute magnitude of galaxy $j$ derived from the redshift $z_{j}$ (taken as a distance proxy within an assumed background cosmology)
and $\overline{M}$ is the mean over all galaxies in the sample. Since $\overline{M}$ depends on $z$,\footnote{Here we assume a uniform galaxy sample where
luminosities at a fixed redshift are drawn from a single distribution. However, additional dependencies on different galaxy types (e.g., spirals and ellipticals) can
be easily accounted for by considering appropriate subsamples \citep{Nusser2011}.} we envisage two possibilities of determining it from the data.
The first one is to divide the sample into multiple redshift bins which are then used to compute individual estimates of $\overline{M}$. Alternatively, $\overline{M}$
can be obtained through the global luminosity function at $z_{\rm eff}$ if one accounts for an effective luminosity evolution term in the distance-magnitude relation.

In what follows, we shall restrict ourselves to galaxy redshifts $z < 0.4$--$0.5$ such that any magnification effects in $\Delta M$ due to gravitational lensing may
be safely neglected \citep[e.g.,][]{Yoo2009}. In this case, the large-scale luminosity modulation signal separates to lowest order in perturbation theory into two main
contributions. The first one traces the radial peculiar velocity field whereas the second component arises from environmental dependencies of the galaxy luminosity
distribution and reflects the fact that more luminous galaxies tend to populate higher-density regions \citep{Mo2004}. Additional scatter due to the natural spread in
galaxy luminosities is not expected to be correlated with any of these components and will only contribute Poissonian noise.

It is useful to express luminosity modulations in terms of an associated radial velocity $u_{M,j}$,
\begin{equation}
u_{M,j} = u_{j} + u_{{\rm env},j} + \epsilon_{j},
\label{eq:212}
\end{equation}
where $u_{j}$ and $u_{{\rm env},j}$ are the radial peculiar velocity component and the environmental luminosity dependence, respectively, and $\epsilon_{j}$ is assumed
to be an uncorrelated error corresponding to the intrinsic scatter of galaxy magnitudes.
To translate a magnitude shift, $\Delta M_{j}$, into its associated radial peculiar velocity, $u_{M,j}$, we will adopt a linear relation for simplicity. The relation is
obtained from expanding the distance modulus ${\rm DM}$ at $z_{j}$ in terms of a small redshift perturbation, where
\begin{equation}
{\rm DM}(z) = 25 + 5\log_{\rm 10}\lbrack D_{L}(z)/{\rm Mpc}\rbrack
\label{eq:214}
\end{equation}
and $D_{L}$ denotes the cosmological luminosity distance in units of Mpc. Since the variance of $u_{M,j}$ is dominated by the error $\epsilon_{j}$, we approximately have
\begin{equation}
\sigma^{2}_{u} \approx \langle\epsilon^{2}\rangle = \frac{1}{N}\sum_{j}\sigma^{2}_{u,j},
\label{eq:213}
\end{equation}
where $N$ is the number of available galaxies in the sample. Again, this quantity (used to quantify shot noise contributions) may be estimated from the data through
the observed scatter of magnitudes, $\sigma^{2}_{M}=\sum_{j}\sigma^{2}_{M,j}/N$, where $\sigma^{2}_{M,j}=\Delta M^{2}_{j}$ \citep[cf.][]{Nusser2013}.
 
The above will form the basis for our study of luminosity-density correlations in the next sections. Despite its simplicity, our approach will suffice to get a first
assessment of how well measurements of such correlations on scales $\gtrsim 10h^{-1}$ Mpc ($h$ is the dimensionless Hubble constant) can be used as a cosmological probe.

\subsection{Luminosity-density correlations}
\label{section22}
Focusing on large scales, where linear perturbation theory is applicable, the density contrast of galaxies, $\delta_{g}$, is related to that of matter, $\delta$, through
the linear local bias factor $b$, i.e. $\delta_{g}=b\delta$. Expressing all quantities in terms of their Fourier transforms and neglecting the effects of RSDs and shot
noise for the moment, we will be interested in correlations of the form
\begin{equation}
\langle u_{M}\delta_{g}^{*}\rangle = VP_{M\delta} = \langle u\delta_{g}^{*}\rangle + \langle u_{\rm env}\delta_{g}^{*}\rangle
\label{eq:221}
\end{equation} 
and
\begin{equation}
\begin{split}
\langle u_{M}u_{M}^{*}\rangle &= VP_{MM}\\
&= \langle uu^{*}\rangle + \langle uu_{\rm env}^{*}\rangle + \langle u^{*}u_{\rm env}\rangle + \langle u_{\rm env}u_{\rm env}^{*}\rangle,
\end{split}
\label{eq:222}
\end{equation}
where $P_{M\delta}$ and $P_{MM}$ denote the corresponding power spectra, and $V$ is the sample volume. The large-scale dependence of galaxy luminosities on
their environment is strongly supported by several studies \citep[e.g.,][]{Balogh2001, Mo2004, Croton2005, Park2007, Merluzzi2010, Faltenbacher2010, Zehavi2011},
pointing towards the large-scale environment's overdensity as the most important factor on scales of a few Mpc. Lacking observational constraints on the large
scales relevant to this work, we will assume that the observed environmental dependence can be extrapolated to scales of $10$--$100h^{-1}$ Mpc.

The distinct angular dependence of contributions due to the peculiar velocity field allows for separating these two effects, making it possible to extract them
from the measured luminosity-based signal. 
To illustrate this point, let us simplify the problem by considering the above correlations in the distant observer approximation. Further, we will model effects
due to environment by assuming that $u_{\rm env}$ only depends on the local density contrast, irrespective of the adopted smoothing scale \citep{Mo2004, Faltenbacher2010,
Yan2013}. Setting $u_{\rm env}=\alpha\delta$ and using the linear velocity-density relation
\begin{equation}
u = -i\mu\frac{aHf}{k}\delta,
\label{eq:223}
\end{equation}
where $a$ denotes the cosmic scale factor, $H$ is the Hubble constant, $f={\rm d}\log{D}/{\rm d}\log{a}$ is the linear growth rate of density perturbations, and $\mu$ is
the cosine of the angle between the wave vector $\bm{k}$ and the line of sight ($\mu=k_{3}/k$, $k_{3}$ denotes the line-of-sight component, and $k=\lvert\bm{k}\lvert$),
the spectra in Eqs. \eqref{eq:221} and \eqref{eq:222} take the form
\begin{equation}
\begin{split}
P_{M\delta} &= \mu br_{g}\tilde{P}_{u\delta} + \alpha bP_{\delta\delta},\\
P_{MM} &= \mu^{2}\tilde{P}_{uu} + \alpha^{2}P_{\delta\delta},
\end{split}
\label{eq:224}
\end{equation}
where we have introduced the galaxy correlation coefficient $r_{g}$ and
\begin{equation}
\tilde{P}_{u\delta} = i\frac{aHf}{k}P_{\delta\delta},\qquad
\tilde{P}_{uu} = \left (\frac{aHf}{k}\right )^{2}P_{\delta\delta}.
\label{eq:225}
\end{equation}
Note that the coefficient $r_{g}$ may take values smaller than unity in the case of stochastic biasing \citep{dh99} and that the velocity-density spectrum is
purely imaginary, which leads to a vanishing cross term in the expression for $P_{MM}$.

Expanding the spectra in Eq. \eqref{eq:224} in terms of multipoles, one finds that the monopole and dipole of $P_{M\delta}$ are given by
\begin{equation}
P_{M\delta}^{(0)} = \alpha bP_{\delta\delta},\qquad P_{M\delta}^{(1)} = br_{g}\tilde{P}_{u\delta},
\label{eq:226}
\end{equation}
and the monopole of $P_{MM}$ is
\begin{equation}
P_{MM}^{(0)} = \frac{1}{3}\tilde{P}_{uu} + \alpha^{2}P_{\delta\delta}.
\label{eq:227}
\end{equation}
Equation \eqref{eq:226} shows that the dipole of $P_{M\delta}$ probes cosmological velocity-density correlations whereas the monopole isolates the effect due to
environment. The monopole of $P_{MM}$ traces a combination of both contributions and does not depend on $b$ and $r_{g}$. In principle, it is also possible to
extract the velocity contribution from $P_{MM}$ by considering its quadrupole. As will become clear below, however, the prospects of measuring its signal are
rather dim, and we choose not to discuss this possibility further.

To quantify how well the monopole and dipole of $P_{M\delta}$ could be constrained from observations, we will assume Gaussian statistics and adopt a standard
count-in-cells approach to estimate statistical uncertainties \citep[e.g.,][]{Peeb80, Smith2009}. Because the calculation closely follows the multipole analysis
presented in \cite{Feix2013}, we will only sketch the derivation here. We start from discrete representations of the fields in Fourier space, i.e.
\begin{equation}
u_{M}\left (\bm{k}\right ) = \frac{1}{\overline{n}V}\sum\limits_{\gamma}u_{M,\gamma}n_{\gamma}\exp{\left (i\bm{k}\bm{r}_{\gamma}\right )}
\label{eq:228a}
\end{equation}
and
\begin{equation}
\delta\left (\bm{k}\right ) = \frac{1}{\overline{n}V}\sum\limits_{\gamma}\left (n_{\gamma} - \left\langle n_{\gamma}\right\rangle\right )
\exp{\left (i\bm{k}\bm{r}_{\gamma}\right )},
\label{eq:228b}
\end{equation}
where the index $\gamma$ runs over infinitesimal cells that sample the volume $V$ and contain at most a single object ($n_{\gamma}=0,1$) and $\overline{n}$ is the
mean number density defined by $\langle n_{\gamma}^{2}\rangle = \langle n_{\gamma}\rangle = \overline{n}\delta V_{\gamma}$. The product $u_{M}\delta^{*}$ is then
used to construct estimators of the monopole and the dipole, $\hat{P}_{M\delta}^{(0)}$ and $\hat{P}_{M\delta}^{(1)}$, respectively. The last step consists of determining
the expected variance of these estimators to leading order in $(\overline{n}V)^{-1}$. Assuming $b=r_{g}=1$ for simplicity, we arrive at
\begin{equation}
\begin{split}
\left\lvert\Delta P_{M\delta}^{(0)}\right\rvert^{2} &= \frac{1}{2N_{k}}\int\limits_{-1}^{1}{\rm d}\mu\left\lbrack\left (P_{MM} + \frac{\sigma^{2}_{u}}{\overline{n}}\right )\right.\\
&\phantom{==} \times \left.\left (P_{\delta\delta} + \frac{1}{\overline{n}}\right ) + P_{M\delta}\left (\bm{k}\right)P_{M\delta}\left (-\bm{k}\right )\right\rbrack\\
&= \frac{1}{N_{k}}\left\lbrack\left (\frac{1}{3}\tilde{P}_{uu} + \alpha^{2}P_{\delta\delta} + \frac{\sigma^{2}_{u}}{\overline{n}}\right )\right.\\
&\phantom{==} \times \left.\left (P_{\delta\delta} + \frac{1}{\overline{n}}\right ) - \frac{1}{3}\tilde{P}_{u\delta}^{2} + \alpha^{2}P_{\delta\delta}^{2}\right\rbrack
\end{split}
\label{eq:229a}
\end{equation}
and
\begin{equation}
\begin{split}
\left\lvert\Delta P_{M\delta}^{(1)}\right\rvert^{2} &= \frac{9}{2N_{k}}\int\limits_{-1}^{1}{\rm d}\mu\mu^{2}\left\lbrack\left (P_{MM} + \frac{\sigma^{2}_{u}}{\overline{n}}\right )\right.\\
&\phantom{==} \times \left.\left (P_{\delta\delta} + \frac{1}{\overline{n}}\right ) - P_{M\delta}\left (\bm{k}\right)P_{M\delta}\left (-\bm{k}\right )\right\rbrack\\
&= \frac{3}{N_{k}}\left\lbrack\left (\frac{3}{5}\tilde{P}_{uu} + \alpha^{2}P_{\delta\delta} + \frac{\sigma^{2}_{u}}{\overline{n}}\right )\right.\\
&\phantom{==} \times \left.\left (P_{\delta\delta} + \frac{1}{\overline{n}}\right ) + \frac{3}{5}\tilde{P}_{u\delta}^{2} - \alpha^{2}P_{\delta\delta}^{2}\right\rbrack,
\end{split}
\label{eq:229b}
\end{equation}
where
\begin{equation}
N_{k} = \frac{1}{2\pi^{2}}k^{2}\Delta kV_{\rm eff}
\end{equation}
is the number of available modes for a redshift survey with effective volume $V_{\rm eff}$ and $k$-binning $\Delta k$. A similar expression can be
obtained for the monopole of $P_{MM}$.

In practice, galaxies are observed in redshift space, and thus a realistic analysis should consider correlations with the redshift-space density contrast
$\delta^{s}=(1+f\mu^{2})\delta$, where it is again assumed that $b=r_{g}=1$. An obvious effect caused by RSDs is a boost in the multipole signals,
\begin{equation}
P_{M\delta^{s}}^{(0)} = \alpha\left (1+\frac{1}{3}f\right )P_{\delta\delta},\qquad
P_{M\delta^{s}}^{(1)} = \left (1+\frac{3}{5}f\right )\tilde{P}_{u\delta}.
\end{equation}
However, there is also a subtlety related to using the distant observer approximation. As discussed in \cite{Nusser2017}, the velocity-density correlation
function in redshift-space, $\xi^{s}_{u\delta}$, contains a term proportional to $\xi_{uu}/r$, where $\xi_{uu}$ is the velocity autocorrelation function.
This yields additional contributions to both the monopole and the dipole which are missing in our present treatment. Since these contributions decrease
with $1/r$, they are mostly relevant to local galaxy data. For example, choosing $r=50h^{-1}$ Mpc already results in a negligible correction to the dipole
of $P_{M\delta}$. For even larger radii $r\gtrsim 250h^{-1}$ Mpc, we thus do not expect any significant deviations from our findings, suggesting that the
expressions derived for $P_{M\delta}$ and $P_{MM}$ may be safely adopted to estimate statistical errors associated with such correlation measurements.

\subsection{Expectations for SDSS}
\label{section23}
As a first example, let us consider a galaxy catalogue comparable to the Sloan Digital Sky Survey (SDSS) main galaxy sample \citep{York2000, SDSSDR13} at
an effective depth $r_{\rm eff}\approx 300h^{-1}$ Mpc. To estimate the statistical uncertainty of luminosity-density correlation measurements, we assume
that galaxies approximately reside at a fixed redshift $z=0.1$. The effect of environment is modelled by resorting to the linear relation $\Delta M_{\rm env}=0.2\delta$
that follows from observations in the visible band \citep{Croton2005, Mercurio2012} and approximately gives $\alpha/c\approx 8.5\times 10^{-3}$, where $c$ is
the speed of light. From SDSS galaxies in the $r$-band, one roughly estimates $\sigma_{M}\approx 0.5$ which translates into $\sigma_{u}\approx 0.02c$ at
the given redshift (see section \ref{section21}). For the actual computation, we will adopt a flat $\Lambda$CDM cosmology with fixed parameters taken from
\cite{Calabrese2013} and the parametrised linear matter power spectrum of \cite{EH98} smoothed on a scale of $10h^{-1}$ Mpc. Further, we set $\overline{n}=0.01
(h^{-1}$ Mpc)$^{-3}$ and the effective volume $V_{\rm eff}=0.2(h^{-1}$ Gpc)$^{3}$ \citep{Percival2010}, and assume a binning width $\Delta k=0.01$ Mpc$^{-1}$.

\begin{figure}
\includegraphics[width=0.95\linewidth]{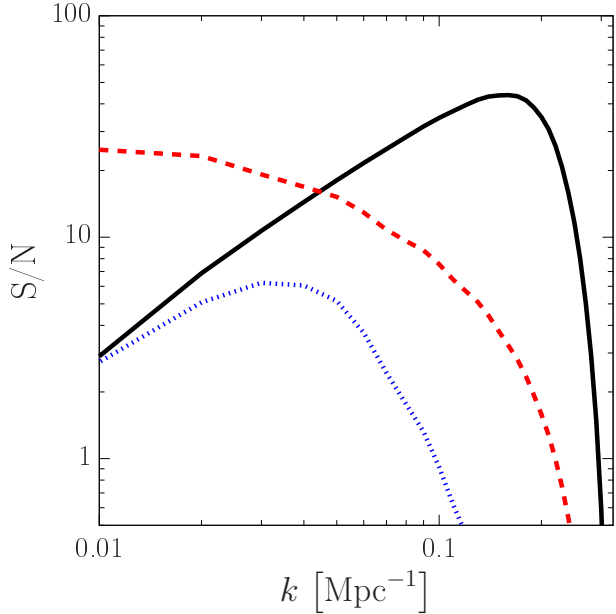}
\caption{Signal-to-noise ratios (S/N) as a function of $k$ for the monopole (black solid line) and the dipole (red dashed line) of $P_{M\delta}$ for
an SDSS-like galaxy redshift survey. The estimates assume a binning width $\Delta k=0.01$ Mpc$^{-1}$ and are based on Eqs. \eqref{eq:226}, \eqref{eq:229a}, and
\eqref{eq:229b}, adopting a linear $\Lambda$CDM power spectrum smoothed on a scale of $10h^{-1}$ Mpc and the parameters given in the text. In addition, the figure
shows S/N for the monopole of $P_{MM}$ (blue dotted line) for the case of no environmental dependence in the luminosity distribution, i.e. $\alpha=0$.}
\label{fig1}
\end{figure}

The resulting signal-to-noise ratios (S/N) for the monopole and dipole of $P_{M\delta}$ are presented in Fig. \ref{fig1}. Our analysis indicates that both
signals can be constrained with relatively high ${\rm S/N}\gtrsim 10$ on scales of about 10--100 Mpc. In reality, questionable parts of the data and additional
cuts might yield slightly lower values of these ratios, but the signals should remain well detectable. As for luminosity autocorrelations, the reduced signal
amplitude will probably not allow for a reliable measurement of the velocity signal in current surveys. This is illustrated in the figure by the blue dotted
line which shows the expected S/N for the monopole of $P_{MM}$ for $\alpha=0$. However, tracing a combination of the velocity signal and environmental effects,
estimates of this quantity will still be useful to obtain upper limits.

A general concern of luminosity-based techniques is the photometric accuracy of the data. Uncertainties in the photometric calibration might propagate into
systematic errors that exhibit a coherent structure on large scales and could mimic spurious flows, leading to biases in the measurements. The precise nature
of such systematics depends on the used instruments and survey strategy. Already in current survey such as SDSS, their impact is rather small and amounts to
relative deviations at the level of 1\% or less \citep{Finkbeiner2016}. It is expected that these will be further reduced in future surveys. Measurements of luminosity-density
correlations should be particularly robust to the presence of systematic photometric errors because they are unlikely to be correlated with the underlying
density field \citep{Nusser2012, Feix2017}. In principle, also uncertainties in the radial selection function and the modelling of $K$-corrections \citep[e.g.,]
[]{Blanton2007} and luminosity evolution are potentially problematic, but we believe that any resulting effects should be mitigated due to the fact that
luminosity modulations are derived relative to an \emph{observed} mean distribution. Further complications may arise from partial sky coverage causing a mixing
of different multipoles. Clearly, these various issues deserve further study and can be investigated in controlled experiments based on appropriate galaxy mock
catalogues or through consistency checks that are directly applied to the data. For example, one possibility would be to consider the analysis for subsamples
defined by different mean redshifts or density environments. Another option is to include comparisons to estimates of the full large-scale peculiar velocity field
that appear remarkably robust against systematic errors of this kind \citep{Feix2015, Feix2017}.

\section{Conclusions}
\label{section3}
Direct measurements of radial peculiar velocities allow for interesting constraints on cosmic velocity-density correlations, but are limited to rather local
volumes due to various observational challenges. Here we propose an alternative route by correlating spatial modulations in the observed galaxy luminosity
distribution of a redshift survey with the underlying density field. For redshifts limited to roughly $z<0.4$--$0.5$, the monopole part of this signal essentially
traces the environmental dependence of galaxy luminosities while the dipole probes large-scale velocity-density correlations. Considering an SDSS-like galaxy
survey at $z=0.1$, we have estimated that both the monopole and the dipole signal could be constrained with relatively high ${\rm S/N}\gtrsim 10$ on scales of
about 10--100 Mpc.

Observational constraints on the environmental dependence of galaxy luminosities provide an important test of galaxy formation models that associate galaxies
with dark matter halos \citep{Mo2004}. These dependencies are usually studied through the galaxy luminosity function in different density environments or by
measuring galaxy clustering properties as a function of luminosity and morphological type \citep{Skibba2006, Zehavi2011, McNaught2014}, indicating that local density is the
most relevant parameter on the large scales considered here. Probing these effects independently with luminosity-density correlations will provide a valuable
addition. Such measurements could also be helpful to quantify the importance of environmental dependencies for previous applications of luminosity-based methods
that have been used to constrain bulk flows and the linear growth rate of density perturbations \citep{Nusser2011, Nusser2012, Branchini2012, Feix2014, Feix2015,
Feix2017}.  

Similarly, reliably extracting the peculiar velocity signal from these correlations beyond redshifts $z\sim 0$ is of great importance for constraining the
underlying cosmological model which is sensitive to the nature of dark energy and gravity. Velocity-density correlations play a role in modelling the effects of
RSDs on the density autocorrelation function that is determined from redshift surveys \citep{Koda2014, Okumura2014, Sugiyama2016}. The analysis of luminosity-density
correlations can provide direct observational constraints on this signal and should prove useful in this context. Tracing velocity-density correlations, the dipole
of the luminosity-density correlation function is sensitive to the cosmic growth rate and can be used to estimate this quantity as well \citep{Nusser2017, Adams2017}.
To this end, templates of the velocity-density correlation function that are based on simulated mock catalogues or analytic methods \citep[e.g.,][]{Bartelmann2016}
might be expedient tools. 

For galaxy data at larger redshifts $z\gtrsim 0.5$, one needs to account for magnification due to gravitational lensing caused by the large-scale structure
along the line of sight to individual galaxies. Similar to peculiar velocities, the lensing magnification changes the apparent brightness of galaxies and thus
contributes to the overall budget of observed luminosity modulations. This effect introduces additional anisotropy in the correlations and is most prominent for
separations close to the line of sight, but typically starts to become relevant on very large scales $\gtrsim 100$ Mpc \citep{Hui2007, Hui2008, Yoo2010}.
Since the magnification may be modelled as part of the cosmological signal, it should also be worthwhile to consider the analysis of angular luminosity-density
correlations, which is particularly relevant to photometric redshift catalogues \citep{LSST2012, Abbott2016}. This approach may be considered an extension of
\cite{Nusser2013} who proposed a measurement of the angular luminosity modulation power spectrum to probe the signal amplitudes of both peculiar velocities and
the gravitational lensing magnification out to $z\sim 1$.

We conclude that measurements of luminosity-density correlations constitute an interesting target for probing cosmic physics with current
\citep{SDSSDR13} and next-generation redshift surveys featuring accurate photometry and large numbers of galaxies \citep[e.g.,][]{euclid2011, daCunha2017}.

\section*{Acknowledgements}
The author thanks M. Bartelmann, M. Bilicki, W. A. Hellwing, A. Nusser, and B. M. Sch\"{a}fer for comments on the manuscript.
This research was supported by the Excellence Initiative of the German Federal and State Governments at Heidelberg University.




\bibliographystyle{mnras}
\bibliography{lumcorr}








\bsp	
\label{lastpage}
\end{document}